\documentclass[aps,prl,preprint,groupedaddress,nofootinbib]{revtex4-1}

\usepackage{slashed}
\newcommand{\re}{\mathop{\mathrm{Re}}\nolimits}
\newcommand{\im}{\mathop{\mathrm{Im}}\nolimits}

\begin{document}


\title{
\vskip-3cm{\baselineskip14pt
\centerline{\normalsize\rm DESY 14-140\hfill ISSN 0418-9833}
\centerline{\normalsize\rm MPP-2014-323\hfill}
\centerline{\normalsize\rm August 2014\hfill}}
\vskip1.5cm
Considerations concerning the generalization of the Dirac equations to
unstable fermions}


\author{Bernd A. Kniehl}
\email[]{kniehl@desy.de}
\altaffiliation{Permanent address: II. Institut f\"ur Theoretische Physik,
Universit\"at Hamburg, Luruper Chaussee 149, 22761 Hamburg, Germany}
\author{Alberto Sirlin}
\email[]{alberto.sirlin@nyu.edu}
\altaffiliation{Permanent address: Department of Physics, New York University,
4~Washington Place, New York, New York~10003, USA}
\affiliation{Max-Planck-Institut f\"ur Physik (Werner-Heisenberg-Institut),
F\"ohringer Ring 6, 80805 Munich, Germany}


\date{\today}

\begin{abstract}
We discuss the generalization of the Dirac equations and spinors in momentum
space to free unstable spin-$1/2$ fermions taking into account the fundamental
requirement of Lorentz covariance.
We derive the generalized adjoint Dirac equations and spinors, and explain the
very simple relation that exists, in our formulation, between the unstable and
stable cases.
As an application of the generalized spinors, we evaluate the probability
density.
We also discuss the behavior of the generalized Dirac equations under time
reversal.
\end{abstract}


\maketitle


The necessity to generalize the Dirac equations and spinors in momentum space
to free unstable spin-$1/2$ particles has recently been recognized in
connection with the wave-function renormalizations of mixed systems of Dirac
\cite{Kniehl:2008cj,Kniehl:2014dra} and Majorana fermions
\cite{Kniehl:2014gfa}.
In this report, we discuss their construction when the fundamental requirement
of Lorentz covariance is taken into account.
We also derive the generalized adjoint Dirac equations and spinors, and explain
the very simple relation that exists, in our formulation, between the
generalized Dirac equations and spinors and the corresponding expressions for
stable fermions.\footnote{%
For brevity, spin-$1/2$ particles are henceforth called fermions.}
We illustrate the application of the generalized spinors by evaluating the
probability density.   

Defining the complex mass $M$ of the unstable fermion as the zero of the
inverse propagator, a frequently used parametrization is \cite{Smith:1996xz}
\begin{equation}
M=m-i\frac{\Gamma}{2},
\label{eq:pole}
\end{equation}
where $m$ and $\Gamma$ are its mass and width, respectively.

We define the four-momentum of the unstable fermion according to
\begin{equation}
p^0=M\gamma c,\qquad
\vec{p}=M\gamma\vec{v},
\label{eq:p}
\end{equation}
where $\gamma=(1-\vec{v}^{\,2}/c^2)^{-1/2}$ and $\vec{v}$ is the particle's
velocity in the chosen inertial frame.
Since Eq.~(\ref{eq:p}) differs from the expressions of special relativity
for stable fermions by only the constant factor $M/m$, $p^\mu=(p^0,\vec{p}\,)$
transforms as a four-vector.
Equation~(\ref{eq:p}) can also be written as
\begin{equation}
p^\mu=Mu^\mu,
\label{eq:u}
\end{equation}
where $u^\mu=\gamma(c,\vec{v}\,)$ is the four-velocity.
From Eqs.~(\ref{eq:p}) and (\ref{eq:u}) one finds the basic relation\footnote{%
In this paper, we adopt the notations and conventions of
Refs.~\cite{bjorken,mandl}.}
\begin{equation}
p_\mu p^\mu={p^0}^2-\vec{p}^{\,2}=M^2c^2.
\label{eq:ps}
\end{equation}
Evaluating the real and imaginary parts of $p^0$ from Eq.~(\ref{eq:ps}) and
using the relation $\im\vec{p}^{\,2}=-[m\Gamma/(m^2-\Gamma^2/4)]\re\vec{p}^{\,2}$
that follows from Eqs.~(\ref{eq:pole}) and (\ref{eq:p}), one can express $p^0$
in terms of $\re\vec{p}^{\,2}$, $m$, and $\Gamma$ as
\begin{equation}
p^0=
M\left[\frac{\re\vec{p}^{\,2}+(m^2-\Gamma^2/4)c^2}{m^2-\Gamma^2/4}\right]^{1/2}.
\label{eq:p0}
\end{equation}
In the limit $\Gamma\to0$, $\vec{p}^{\,2}$ is real, $p^0=E/c$, and
Eq.~(\ref{eq:p0}) becomes
\begin{equation}
E=(\vec{p}^{\,2}c^2+m^2c^4)^{1/2},
\end{equation}
the well-known energy-momentum relation for stable particles.
In the rest frame, $\vec{p}=\vec{0}$ and Eq.~(\ref{eq:p0}) reduces to $p^0=Mc$,
in agreement with Eq.~(\ref{eq:p}) when $\vec{v}=\vec{0}$.

In momentum space, the generalizations of the Dirac equations to free unstable
fermions are
\begin{equation}
(\slashed{p}-Mc)u_r(\vec{p}\,)=0,\qquad
(\slashed{p}+Mc)v_r(\vec{p}\,)=0,
\label{eq:dirac}
\end{equation}
where $\slashed{p}=p_\mu\gamma^\mu$ and $r=1,2$ labels the two independent
solutions.
Recalling Eq.~(\ref{eq:p}), we note that Eq.~(\ref{eq:dirac}) can be derived by
multiplying the corresponding Dirac equations for stable fermions by $M/m$.
The four independent solutions can be written explicitly in the form
\begin{equation}
u_r(\vec{p}\,)=\left(\frac{p^0+Mc}{2Mc}\right)^{1/2}\left(
\begin{array}{c}
\chi_r \\
\frac{\vec{\sigma}\cdot\vec{p}}{p^0+Mc}\chi_r
\end{array}\right),
\qquad
v_r(\vec{p}\,)=\left(\frac{p^0+Mc}{2Mc}\right)^{1/2}\left(
\begin{array}{c}
\frac{\vec{\sigma}\cdot\vec{p}}{p^0+Mc}\chi_r^\prime \\
\chi_r^\prime
\end{array}\right),
\label{eq:spinor}
\end{equation}
where $\sigma^i$ are the Pauli matrices and $\chi_r$ and $\chi_r^\prime$ are
two-dimensional constant and orthogonal spinors frequently chosen as
\begin{equation}
\chi_1=\chi_2^\prime=\left(
\begin{array}{c}
1 \\
0
\end{array}\right),
\qquad
\chi_2=\chi_1^\prime=\left(
\begin{array}{c}
0 \\
1
\end{array}\right).
\label{eq:chi}
\end{equation}
With this choice, $u_r(\vec{p}\,)$ and $v_r(\vec{p}\,)$ are eigenstates of the
$z$ component of spin in the rest frame of the fermion, with eigenvalue
$+\hbar/2$ (spin up) for $u_1(\vec{p}\,)$ and $v_2(\vec{p}\,)$ and $-\hbar/2$
(spin down) for $u_2(\vec{p}\,)$ and $v_1(\vec{p}\,)$.

Including the space-time dependencies, the plane-wave solutions associated with
the spinors $u_r(\vec{p}\,)$ and $v_r(\vec{p}\,)$ are
$u_r(\vec{p}\,)\exp(-ip\cdot x)$ and $v_r(\vec{p}\,)\exp(ip\cdot x)$,
respectively.
Using Eqs.~(\ref{eq:pole})--(\ref{eq:u}), we have
\begin{equation}
e^{-ip\cdot x}=e^{-imu\cdot x}e^{-(\Gamma/2)u\cdot x}.
\label{eq:exp}
\end{equation}
The first factor on the right-hand side of Eq.~(\ref{eq:exp}) is the space-time
dependence in the stable case, while the second factor reflects the fact that
the fermion is unstable.
The amplitude $u_r(\vec{p}\,)\exp(-imu\cdot x)$ is a solution of the usual
Dirac equation for stable fermions and is, therefore, time-reversal invariant.
By contrast, the second factor,
$\exp[-\Gamma\gamma(c^2t-\vec{v}\cdot\vec{x})/2]$, is not invariant under the
time-reversal transformation $t\to-t$ and $\vec{v}\to-\vec{v}$.

Another simple way to show that the generalized Dirac equations are not
invariant under time reversal is the following:
we recall that the operator that relates the wave functions at times $t$ and
$t^\prime=-t$ is antiunitary, namely of the form $KU$, where $U$ is a unitary
matrix and $K$ means complex conjugation.
If the Hamiltonian $H(t)$ at time $t$ involves the complex mass $M$, as is the
case in the formulation of the generalized Dirac equations, when $K$ acts on
$H(t)$ it transforms $M\to M^*$.
As a consequence, $H(t^\prime)$ differs from $H(t)$ by the same change
$M\to M^*$, and the proof of time-reversal invariance, explained for instance
in Ref.~\cite{bjorken}, breaks down.

The Hermitian adjoints of Eq.~(\ref{eq:dirac}) are
\begin{equation}
\bar{u}_r(\vec{p}\,)(\slashed{p}^*-M^*c)=0,\qquad
\bar{v}_r(\vec{p}\,)(\slashed{p}^*+M^*c)=0,
\label{eq:diracc}
\end{equation}
where
\begin{equation}
\bar{u}_r(\vec{p}\,)=u_r^\dagger(\vec{p}\,)\gamma^0,\qquad
\bar{v}_r(\vec{p}\,)=v_r^\dagger(\vec{p}\,)\gamma^0
\end{equation}
are the usual adjoint spinors and $\slashed{p}^*=p_\mu^*\gamma^\mu$.
At first sight, the presence of the complex conjugates $p_\mu^*$ and $M^*$ seems
to complicate matters.
However, we note from Eq.~(\ref{eq:u}) that $p^\mu/M=u^\mu$ is real.
Therefore, we have the important relation
\begin{equation}
\left(\frac{p^\mu}{M}\right)^*=\frac{p^\mu}{M}.
\label{eq:ratio}
\end{equation}
Inserting $p_\mu^*=(M^*/M)p_\mu$ and multiplying by $M/M^*$,
Eq.~(\ref{eq:diracc}) becomes
\begin{equation}
\bar{u}_r(\vec{p}\,)(\slashed{p}-Mc)=0,\qquad
\bar{v}_r(\vec{p}\,)(\slashed{p}+Mc)=0.
\label{eq:diraca}
\end{equation}
The four generalized Dirac equations shown in Eqs.~(\ref{eq:dirac}) and
(\ref{eq:diraca}) were postulated in Ref.~\cite{Kniehl:2014dra} without
applying Eq.~(\ref{eq:ratio}) and with a different interpretation of the
adjoint spinors $\bar{u}_r(\vec{p}\,)$ and $\bar{v}_r(\vec{p}\,)$.
In the case of unstable fermions, Eqs.~(\ref{eq:dirac}) and (\ref{eq:diraca})
play an important role in the implementation of the
Aoki-Hioki-Kawabe-Konuma-Muta (AHKKM) \cite{Aoki:1982ed} renormalization
conditions in general theories with intergeneration mixing, as pointed out
in Refs.~\cite{Kniehl:2008cj,Kniehl:2014dra,Kniehl:2014gfa}.

Since $p^\mu$ transforms as a four-vector, the proof of Lorentz covariance of
Eq.~(\ref{eq:dirac}) follows the same steps as the proof of the Lorentz
covariance of the Dirac equation in coordinate space (see, for example,
chapter~2 in Ref.~\cite{bjorken}).
Specifically, if $p^\mu$ and $u_r(\vec{p}\,)$ are the four-momentum and the
spinor in the Lorentz frame $O$ and $p^{\prime\mu}$ and
$u_r^\prime(\vec{p}^{\,\prime})$ are those in the Lorenz frame $O^\prime$, one
expresses, for example, the first equality in Eq.~(\ref{eq:dirac}) in terms of
the $O^\prime$ variables by means of the relations
$p_\mu=a_{\phantom{\nu}\mu}^\nu p_\nu^\prime$, where $a_{\phantom{\nu}\mu}^\nu$ are the
coefficients of the Lorentz transformation between the four-vectors $p_\mu$ and
$p_\mu^\prime$, and $u_r(\vec{p}\,)=S^{-1}u_r^\prime(\vec{p}^{\,\prime})$, where $S$
is a matrix that satisfies the relations
$a_{\phantom{\nu}\mu}^\nu S\gamma^\mu S^{-1}=\gamma^\nu$ and
$S^{-1}=\gamma^0S^\dagger\gamma^0$.
Then the first equality in Eq.~(\ref{eq:dirac}) becomes
$(\slashed{p}^\prime-Mc)u_r^\prime(\vec{p}^{\,\prime})=0$, which demonstrates its
Lorentz covariance.
Carrying out the Hermitian conjugation of the $O^\prime$ Dirac equation and
using Eq.~(\ref{eq:ratio}), one finds
$\bar{u}_r^\prime(\vec{p}^{\,\prime})(\slashed{p}^\prime-Mc)=0$, which shows the
Lorentz covariance of the corresponding adjoint Dirac equation,
Eq.~(\ref{eq:diraca}).

The adjoint spinors with respect to $u_r(\vec{p}\,)$ and $v_r(\vec{p}\,)$ in
Eq.~(\ref{eq:spinor}) are
\begin{equation}
\bar{u}_r(\vec{p}\,)=\left(\frac{p^0+Mc}{2Mc}\right)^{1/2}\left(
\chi_r^\dagger,-\chi_r^\dagger\frac{\vec{\sigma}\cdot\vec{p}}{p^0+Mc}\right),
\quad
\bar{v}_r(\vec{p}\,)=\left(\frac{p^0+Mc}{2Mc}\right)^{1/2}\left(
\chi_r^{\prime\dagger}\frac{\vec{\sigma}\cdot\vec{p}}{p^0+Mc},-\chi_r^{\prime\dagger}
\right),
\label{eq:spinora}
\end{equation}
where we have again applied Eq.~(\ref{eq:ratio}) to eliminate $p_\mu^*$ and
$M^*$.
In particular, Eq.~(\ref{eq:ratio}) implies that $[(p^0+Mc)/(2Mc)]^{1/2}$ and
$\vec{p}/(p^0+Mc)$ are real.

It is interesting to note that the four generalized Dirac equations shown in
Eqs.~(\ref{eq:dirac}) and (\ref{eq:diraca}) as well as their spinor solutions
presented in Eqs.~(\ref{eq:spinor}) and (\ref{eq:spinora}) can be obtained from
the corresponding ones for stable fermions, for $\Gamma=0$, by simply
substituting $m\to M$ in their explicit mass dependencies and in the definition
of $p^\mu$ in Eqs.~(\ref{eq:p}) and (\ref{eq:u}).

Since $M$ cancels in the ratios $(p^0+Mc)/(2Mc)$ and $\vec{p}/(p^0+Mc)$, these
factors are the same as in the $\Gamma=0$ case.
It hence follows that the spinor solutions in Eqs.~(\ref{eq:spinor}) and
(\ref{eq:spinora}) satisfy the same normalization and completeness relations as
in the case of stable fermions, which are given, for example, by Eqs.~(A.29)
and (A.30) in Ref.~\cite{mandl}.
In particular,
\begin{eqnarray}
\bar{u}_r(\vec{p}\,)u_s(\vec{p}\,)&=&-\bar{v}_r(\vec{p}\,)v_s(\vec{p}\,)
=\delta_{rs},
\nonumber\\
\bar{u}_r(\vec{p}\,)v_s(\vec{p}\,)&=&-\bar{v}_r(\vec{p}\,)u_s(\vec{p}\,)=0.
\end{eqnarray}

In some applications, the two-component spinors $\chi_r$ and $\chi_r^\prime$ in
Eq.~(\ref{eq:chi}) are replaced by helicity eigenstates $\phi_s$ and
$\phi_s^\prime$, respectively.
Because, in the case of unstable particles, $\vec{p}$ is complex [cf.\
Eqs.~(\ref{eq:p}) and (\ref{eq:u})], the usual helicity projection,
\begin{equation}
\frac{\vec{p}}{|\vec{p}\,|}\cdot\frac{\vec{\sigma}}{2}\phi_s=s\phi_s,
\qquad
\frac{\vec{p}}{|\vec{p}\,|}\cdot\frac{\vec{\sigma}}{2}\phi_s^\prime
=-s\phi_s^\prime,
\label{eq:hel}
\end{equation}
where $|\vec{p}\,|=(\vec{p}^{\,*}\cdot\vec{p}\,)^{1/2}$, leads to complex
eigenvalues $s=\pm(\vec{p}^{\,2})^{1/2}/(2|\vec{p}\,|)$, as can be checked by
applying the helicity projection operator twice.
A consistent alternative is to define the helicity projection operator in
Eq.~(\ref{eq:hel}) according to $\vec{u}/|\vec{u}|\cdot\vec{\sigma}/2$, where
$\vec{u}=\vec{p}/M$ are the spatial components of the four-velocity $u$ given
below Eq.~(\ref{eq:u}), or, equivalently,
$\vec{v}/|\vec{v}|\cdot\vec{\sigma}/2$.

It is also instructive to calculate the probability current using the spinor
solutions in Eqs.~(\ref{eq:spinor}) and (\ref{eq:spinora}).
We find
\begin{equation}
c\bar{u}_r(\vec{p}\,)\gamma^\mu u_r(\vec{p}\,)=
c\bar{v}_r(\vec{p}\,)\gamma^\mu v_r(\vec{p}\,)=\frac{p^\mu}{M}=u^\mu.
\end{equation}
As expected, these currents transform as four-vectors.
In particular, for $\mu=0$ we have
\begin{equation}
cu_r^\dagger(\vec{p}\,)u_r(\vec{p}\,)=
cv_r^\dagger(\vec{p}\,)v_r(\vec{p}\,)=u^0=\gamma c.
\label{eq:pro}
\end{equation}
Since $cu_r^\dagger(\vec{p}\,)u_r(\vec{p}\,)$ and
$cv_r^\dagger(\vec{p}\,)v_r(\vec{p}\,)$ are the probability densities, they
should be real and positive, consistent with Eq.~(\ref{eq:pro}).

Another interesting application is to examine the space-time dependence of the
probability density.
Incorporating the space-time factor $\exp(-ip\cdot x)$ of Eq.~(\ref{eq:exp}) in
$cu_r^\dagger(\vec{p}\,)u_r(\vec{p}\,)$, we find that the latter is multiplied by
an overall factor $\exp[-\Gamma\gamma(c^2t-\vec{v}\cdot\vec{x})]$, which
implies that the probability density for the positive-energy states decreases
exponentially with time, reflecting the fermion's instability.

In summary, (i) we have proposed a simple definition of the (complex)
four-momentum of a free unstable spin-$1/2$ particle [cf.\ Eqs.~(\ref{eq:p})
and (\ref{eq:u})] and shown that it indeed transforms as a four-vector,
(ii) we have derived the generalized Dirac equations in momentum space [cf.\
Eq.~(\ref{eq:dirac})] and found explicit spinor solutions [cf.\
Eq.~(\ref{eq:spinor})],
(iii) we have derived the generalized adjoint Dirac equations [cf.\
Eq.~(\ref{eq:diraca})] and spinors [cf.\ Eq.~(\ref{eq:spinora})] by Hermitian
conjugation of Eqs.~(\ref{eq:dirac}) and (\ref{eq:spinor}), respectively,
taking into account the basic relation in Eq.~(\ref{eq:ratio}),
(iv) using the important fact that $p^\mu$ transforms as a four-vector, we have
shown how the proof of Lorentz covariance carries over to the generalized Dirac
equations and their adjoints,
(v) we have pointed out the very simple relation that exists, in our
formulation, between the generalized Dirac equations and spinors and the
corresponding expressions for stable fermions,
(vi) in particular, we have shown that our spinors and adjoint spinors satisfy
the same normalization and completeness relations as in the case of stable
fermions,
(vii) we have proposed a modified definition of the helicity projection
operator for unstable fermions that leads to real eigenvalues,
(viii) as an illustration, we have applied our spinors and adjoint spinors to
calculate the probability density and found that it satisfies the expected
theoretical properties,
and (ix) we have discussed the behavior of the generalized Dirac equations
under time reversal.
As mentioned after Eq.~(\ref{eq:diraca}), the four generalized Dirac equations
in Eqs.~(\ref{eq:dirac}) and (\ref{eq:diraca}) play an important role in the
implementation of the AHKKM \cite{Aoki:1982ed} renormalization conditions for
unstable fermions in general theories with intergeneration mixing.

\begin{acknowledgments}
We thank the Werner-Heisenberg-Institut for the hospitality extended to us
during a visit when this paper was prepared.
This research was supported in part by the German Research Foundation through
the  Collaborative Research Center No.~676 {\it Particles, Strings and the
Early Universe---The Structure of Matter and Space Time}.
The work of A.S. was supported in part by the National Science Foundation 
through Grant No.\ PHY--1316452. 
\end{acknowledgments}


\begin{thebibliography}{99}

\bibitem{Kniehl:2008cj} 
  B.~A.~Kniehl and A.~Sirlin,
  Phys.\ Rev.\ D {\bf 77}, 116012 (2008)
  [arXiv:0801.0669 [hep-th]];
{\bf 85}, 036007 (2012)
  [arXiv:1201.4333 [hep-ph]];
  B.~A.~Kniehl,
  Phys.\ Rev.\ Lett.\  {\bf 112}, 071603 (2014)
  [arXiv:1308.3140 [hep-ph]].

\bibitem{Kniehl:2014dra} 
  B.~A.~Kniehl,
  Phys.\ Rev.\ D {\bf 89}, 096005 (2014).

\bibitem{Kniehl:2014gfa}
  B.~A.~Kniehl,
  Phys.\ Rev.\ D {\bf 89}, 116010 (2014)
  [arXiv:1404.5908 [hep-th]].

\bibitem{Smith:1996xz}
  M.~C.~Smith and S.~S.~Willenbrock,
  Phys.\ Rev.\ Lett.\ {\bf 79}, 3825 (1997)
  [hep-ph/9612329];
  M.~Passera and A.~Sirlin,
  Phys.\ Rev.\ D {\bf 58}, 113010 (1998)
  [hep-ph/9804309].

\bibitem{bjorken}
J.~D. Bjorken and S.~D. Drell,
{\it Relativistic Quantum Mechanics}
(McGraw-Hill, New York, 1964).

\bibitem{mandl}
F.~Mandl and G.~Shaw,
{\it Quantum Field Theory}
(Wiley, Chichester, 1984).

\bibitem{Aoki:1982ed} 
  K.-i.~Aoki, Z.~Hioki, R.~Kawabe, M.~Konuma, and T.~Muta,
  Prog.\ Theor.\ Phys.\ Suppl.\  {\bf 73}, 1 (1982).

\end{thebibliography}

\end{document}